\documentclass[prb,twocolumn,superscriptaddress,showpacs,preprintnumbers,amsmath,amssymb,floats]
{revtex4}

\usepackage{txfonts}
\usepackage{amssymb}
\usepackage{graphicx}

\begin{document}

\title{Electronic Structure of Single Crystalline NdO$_{0.5}$F$_{0.5}$BiS$_{2}$ Studied
by Angle-resolved Photoemission Spectroscopy}

\author{Z. R. Ye}
\affiliation{State Key Laboratory of Functional Materials for Informatics,
Shanghai Institute of Microsystem and Information Technology (SIMIT),
Chinese Academy of Sciences, Shanghai 200050, China}
\affiliation{State Key Laboratory of Surface Physics, Department of
Physics,  and Advanced Materials Laboratory, Fudan University,
Shanghai 200433, China}

\author{H. F. Yang}
\author{D. W. Shen}\email{dwshen@mail.sim.ac.cn}
\affiliation{State Key Laboratory of Functional Materials for Informatics,
Shanghai Institute of Microsystem and Information Technology (SIMIT),
Chinese Academy of Sciences, Shanghai 200050, China}

\author{J. Jiang}
\author{X. H. Niu}
\author{D. L. Feng}
\affiliation{State Key Laboratory of Surface Physics, Department of
Physics,  and Advanced Materials Laboratory, Fudan University,
Shanghai 200433, China}

\author{Y. P. Du}
\author{X. G. Wan}
\author{J. Z. Liu}
\author{X. Y. Zhu}
\author{H. H. Wen}
\affiliation{National
Laboratory of Solid State Microstructures and Department of Physics,
National Center of Microstructures and Quantum Manipulation,
Nanjing University, Nanjing 210093, China}

\author{M. H. Jiang}
\affiliation{State Key Laboratory of Functional Materials for Informatics,
Shanghai Institute of Microsystem and Information Technology (SIMIT),
Chinese Academy of Sciences, Shanghai 200050, China}

\date{\today}

\begin{abstract}

NdO$_{0.5}$F$_{0.5}$BiS$_{2}$ is a new layered superconductor. We have studied the  low-lying electronic structure of a single crystalline NdO$_{0.5}$F$_{0.5}$BiS$_{2}$ superconductor, whose superconducting transition temperature is 4.87K,  with angle-resolved photoemission spectroscopy.  The Fermi surface consists of two small electron pockets around the X point and shows little warping along the $k_z$ direction. Our results demonstrate the multi-band and two-dimensional nature of the electronic structure. The good agreement between the photoemission data and the band calculations gives the renormalization factor of 1, indicating the rather weak electron correlations in this material. Moreover, we found that the actual electron doping level and Fermi surface size are much smaller than what are expected from the nominal composition, which could be largely explained by the bismuth dificiency.  The small Fermi pocket size and the weak electron correlations found here put strong constraints on theory, and suggest  that the BiS$_2$-based superconductors could be conventional BCS superconductors mediated by the electron-phonon coupling.

\end{abstract}


\maketitle


\section{Introduction}

The recent discovery of BiS$_2$-based superconductors has simulated a lot of research interests \cite{BiOSO,LaO,NdO,PrO,CeO,YbO}. Similar to cuprates and iron-based superconductors, the superconductivity is introduced by doping their insulating parent compounds\cite{BiOSO,LaO,NdO}. As a layered superconductor, the superconducting BiS$_{2}$ planes could be sandwiched by various charge reservoir layers, e.g., Bi$_{2}$O$_{2}$(SO$_{4}$)$_{1-x}$, LaO$_{1-x}$F$_{x}$, and NdO$_{1-x}$F$_{x}$.  After much effort of searching for new members in this family \cite{BiOSO,LaO,NdO,PrO,CeO,YbO}, one fundamental question is now pressing for the answer: whether the BiS$_2$-based superconductors are a new family of layered unconventional superconductors or  the conventional BCS superconductors?

In this context,  various scenarios have been proposed\cite{TBT,Awana,RPA,CDW,Xiangang,PhononLi,ZDW}. On the BCS side,  a large electron-phonon (e-ph) coupling constant is estimated based on certain phonon spectra calculations of these materials. As a result, the calculated superconducting transition temperature ($T_C$) of LaO$_{0.5}$F$_{0.5}$BiS$_{2}$ agrees well with the experiments \cite{Xiangang,PhononLi}. This proposal is supported by several magnetic penetration depth measurements on this family of superconductors, which all argued that the BiS$_2$-based superconductors are the conventional s-wave type superconductors with fully developed gaps \cite{penetration,Morenzoni,Putti}. However, a recent neutron scattering work suggested that the e-ph coupling in LaO$_{0.5}$F$_{0.5}$BiS$_{2}$ could be much weaker than  expected \cite{Neutron}. Band calculations found that the Fermi surface of LaO$_{0.5}$F$_{0.5}$BiS$_{2}$ is  quasi-one-dimensional, which would provide a good nesting condition. It was proposed that  spin fluctuations could be enhanced, which might mediate the pairing  in the BiS$_2$-based superconductors\cite{TBT}. In that case, the superconductivity is unconventional and the electron correlations should be important,  as in cuprates and iron-based superconductors. These controversial theoretical predictions  rely on the specific band structure and especially the Fermi surface topology. However, the direct experimental report on the electronic structure of the BiS$_2$-based superconductors is still lacking so far.

In this article, through angle-resolved photoemission spectroscopy (ARPES), we have systematically studied the electronic structure of the single crystalline NdO$_{0.5}$F$_{0.5}$BiS$_{2}$ superconductor with a $T_C$ of 4.87K. There are two electron-like bands around the X point, forming two small rectangle-shaped pockets at the Fermi energy ($E_F$). The Fermi pocket size is found to be much smaller than the band calculations based on the nominal composition, which might be due to the bismuth dificiency. The electronic structure shows weak $k_z$ dependence, indicating its two-dimensional nature. Further comparison between the band calculations and photoemission data suggests that the electron correlation in this compound is negligible. Our results put strong constraints on the current theoretical models of the pairing mechanism in BiS$_{2}$-based superconductors.


\section{Experimental}

\begin{figure}
\includegraphics[width=8.5cm]{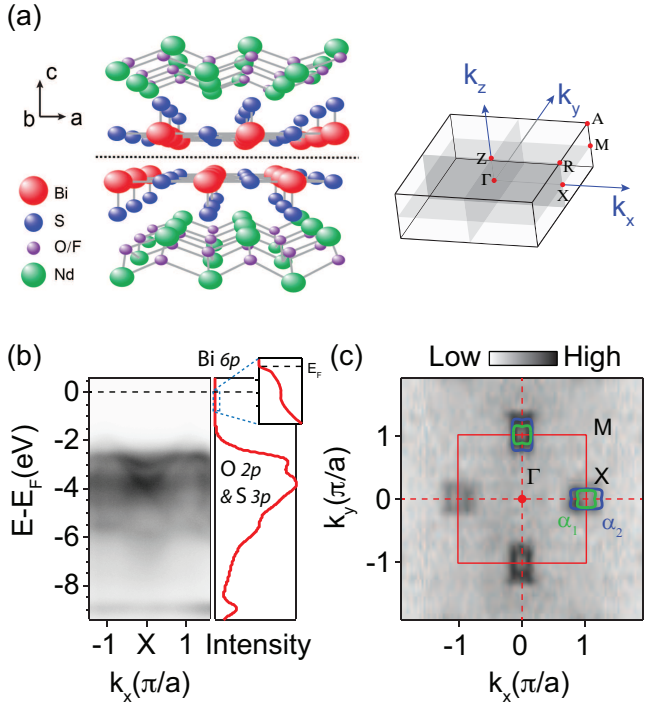}
\caption{(color online) (a) Representative unit cell of NdO$_{1-x}$F$_x$BiS$_{2}$ and its three-dimensional Brillouin zone. The dashed line represents the cleaved plane between the two BiS$_2$ layers. (b) The valence band structure around the X point along the M~-~X direction and corresponding angle-integrated energy distribution curves (EDCs). The inset is the enlarged view of the features near E$_F$. (c) The four-fold symmetrized photoemission intensity map of NdO$_{0.5}$F$_{0.5}$BiS$_{2}$ at $E_F$ over the projected two-dimensional Brillouin zone. The intensity was integrated over a window of ($E_F$-15meV, $E_F$+15meV). The Fermi surface sheets for the $\alpha_1$ and $\alpha_2$ bands are shown by the rectangles with different colors. The data are taken with 100~eV photons.}\label{mapping}
\end{figure}

\begin{figure}[t]
\includegraphics[width=8.5cm]{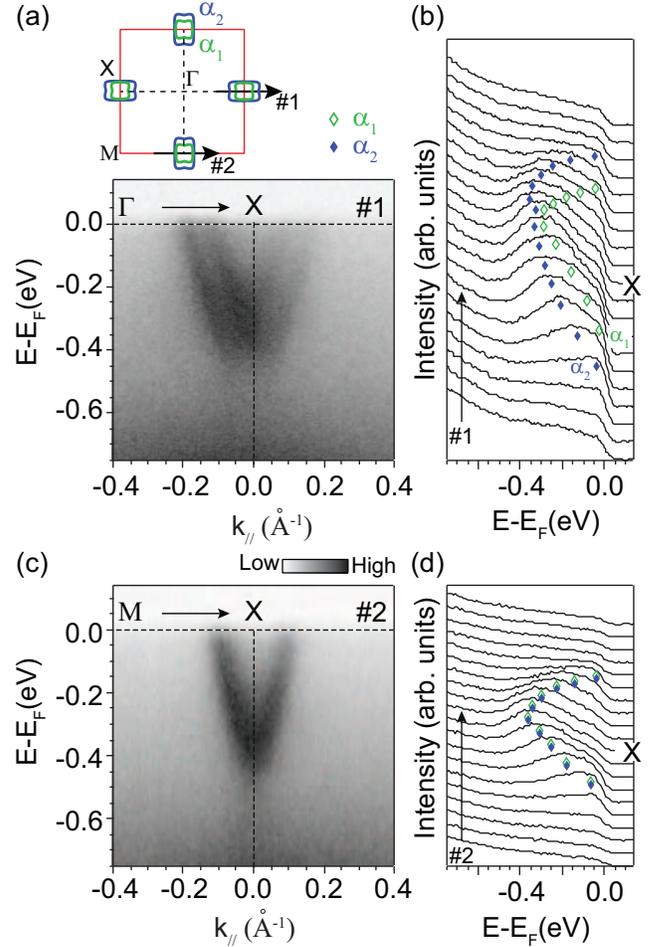}
\caption{ (color online) (a) and (b) are the photoemission intensity and corresponding EDCs along the $\Gamma$~-~X direction. (c) and (d) are the same as (a) and (b), respectively, but taken along M~-~X direction. The data are taken with 49~eV photons, corresponding to the same $k_z$ as that for 100eV photons. }\label{cut}
\end{figure}

High quality NdO$_{0.5}$F$_{0.5}$BiS$_{2}$ single crystals were synthesized with the LiCl/KCl-flux method \cite{WHH}. The actual compositions were determined  to be  NdO$_{0.54}$F$_{0.46}$Bi$_{0.84}$S$_{1.87}$, using  energy-dispersion-spectrum (EDS).  The  bismuth deficiency is rather substantial. The resistivity data (not shown) give a $T_C$ of 4.87K. 
The samples are crystallized in a tetragonal structure with the space group $P4/nmm$, as shown in Fig.~\ref{mapping}(a). There are two possible cleavage planes. One could be between the NdO and BiS$_2$ layers. In this case,  one would, by chance,  encounter  BiS$_2$ or NdO surface, or a surface of partial NdO and BiS$_2$ coverage.  However, while we have cleaved about ten different samples,  we have never observed any different surface. Moreover,  for LaFeAsO \cite{LXYang}, it was found that the LaO surface is a polar one, which contains a metallic surface state with a large Fermi surface. We have not observed such a state, which indicates that the cleavage plane could not be between the NdO/BiS$_2$ layers. On the other hand, the two BiS$_2$ layers are weakly linked by the Van der Waals force, which resembles the two BiO layers  in Bi$_{2}$Sr$_{2}$CaCu$_{2}$O$_{8+\delta}$(Bi2212), so samples should be cleaved much easier between the two BiS$_2$ layers than between the NdO and BiS$_2$ layers [the dashed line in Fig.~\ref{mapping}(a)].  In this case, only one type of surface should be observed, as confirmed by our experiments. Moreover, the cleaved surface in this case is  non-polar without any charge redistribution. The electronic structure measured by ARPES thus should well represent the bulk information.

The ARPES measurements were performed at the SIS beamline of Swiss Light Source (SLS), with a Scienta R4000 electron analyzer. The angular resolution is 0.3 degree, and the overall energy resolution is 15$\thicksim$20~meV depending on the photon energy. The sample surfaces were cleaved $\mathit{in~situ}$ in ultrahigh vacuum with a pressure better than 3$\times$10$^{-11}$ torr. The overall measurements were performed at 15~K, in the normal state of this superconductor. The samples were stable and did not show any sign of degradation during the measurements. The theoretical calculations of the electronic structure are performed within the generalized gradient approximation (GGA), as described elsewhere\cite{Xiangang}.

\section{Results}

\begin{figure*}[t]
\includegraphics[width=17.5cm]{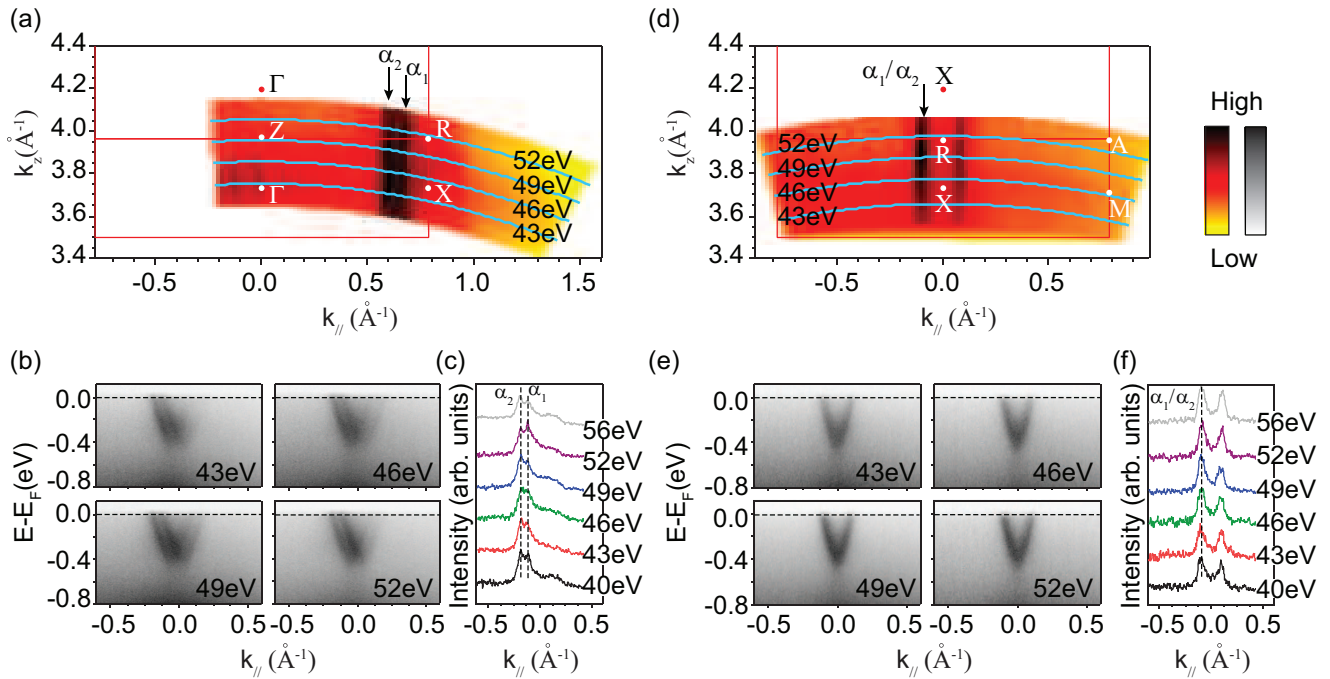}
\caption{(color online)  (a) The photoemission intensity of NdO$_{0.5}$F$_{0.5}$BiS$_{2}$ in the $\Gamma$ZRX plane. The intensity was integrated over a window of ($E_F$-15meV, $E_F$+15meV). Different $k_z$'s were accessed by varying the photon energy as indicated by the solid lines, where an inner potential of 15eV is used to calculate $k_z$. (b) are the photoemission intensities along or parallel to the $\Gamma$~-~X direction, taken at different photon energies. (c) is  the photon energy dependence of the MDCs at $E_F$. (d) is the same as (a), but in the XRAM plane. (e) and (f) are the same as (b) and (c), respectively, but along or parallel to the X~-~M direction.}\label{kz}
\end{figure*}

Figure~\ref{mapping}(b) shows the valence band structure around the X point along the M~-~X high symmetry direction. By comparing with the existing density functional calculations (DFT) \cite{Xiangang}, we can identify that the features between -5eV to -1eV are mainly contributed by the O 2\emph{p} and S 3\emph{p} states, while the small spectral weight near E$_{F}$ can be assigned to the Bi 6\emph{p} state. Figure~\ref{mapping}(c) shows the photoemission intensity map of NdO$_{0.5}$F$_{0.5}$BiS$_{2}$, which is overlaid on the two-dimensional Brillouin zone. Here, the unit of $k{_x}$  and $ k{_y} $ is $ \pi$/$a$, where $a$ is the neighboring Bi-Bi distance in the Bi-S plane. Two rectangle-shaped Fermi pockets were observed around X in the Brillouin zone, which is a direct evidence of the multiband behavior in this compound. This is consistent with previous Hall effect measurements and the theoretical studies as well \cite{CeO,Hall,Hall2,TBT}.

We further examined the low-lying band structure along  \#1 ($\Gamma$~-~X) and \#2 (M~-~X) directions as illustrated in Fig.~\ref{cut}(a). For cut \#1, the photoemission intensity shows two discernible electron-like bands, $\alpha_{1}$  and $\alpha_{2}$. The band dispersion could be resolved by tracking the peaks in the energy distribution curves (EDCs) [Fig.~\ref{cut}(b)]. Both bands disperse to around 300meV below $E_{F}$. The $\alpha_{2}$ band shows a bend-back at the momentum slightly away from the X point, and $\alpha_{2}$ and  $\alpha_{1}$ are degenerate  at the X point. For cut \#2 along M~-~X direction,  only one band can be distinguished within our experimental resolution [Figs.~\ref{cut}(c) and \ref{cut}(d)]. However, considering the four-fold symmetry of the band structure at the X point, there should be two bands along the M~-~X direction, which suggests that $\alpha_{1}$ and $\alpha_{2}$ are degenerate with each other in this direction.

\begin{figure}[t]
\includegraphics[width=8cm]{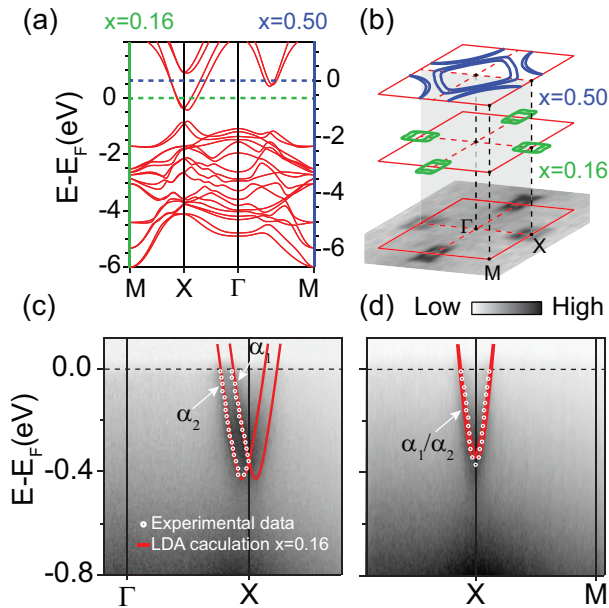}
\caption{ (color online)  (a) The calculated band structures without the spin-orbital coupling for NdO$_{0.84}$F$_{0.16}$BiS$_{2}$ (the left side energy axis) and NdO$_{0.5}$F$_{0.5}$BiS$_{2}$ (the right side one).  (b) The comparison of the Fermi surface between the photoemission intensity map and the DFT for \emph{x}=0.5 and 0.16 fluorine substitution levels. (c) and (d), The comparison of the low-lying band structure near $E_F$ between the photoemission data and the DFT calculations along the $\Gamma$~-~X direction and the X~-~M direction, respectively. The white circles are the peak positions of the MDCs through Lorentzian fitting and the red solid lines are the band dispersions from DFT calculations. The ARPES data are taken with 40~eV photons.}\label{theory}
\end{figure}

To comprehensively understand the electronic structure in the three-dimensional Brillouin zone, we have performed detailed $k_z$ dependent measurements. Figures~\ref{kz}(a) and \ref{kz}(d) show the photoemission intensity maps in the $\Gamma$ZRX and XRAM planes, respectively. Here, we estimated all the $k_z$ values (as illustrated by solid blue lines in the figure) according to the free-electron final-state model\cite{Hufner}, where an inner potential of 15eV was used. The cross sections of both $\alpha_{1}$ and $\alpha_{2}$ Fermi surface sheets show little warping along the $k_z$ direction, indicating the strong two-dimensional nature of the electronic structure. Such a  two-dimensional character is further proved by both the weak variation of the photoemission intensities upon varying the photon energies [Figs.~\ref{kz}(b) and \ref{kz}(e)], and the weak $k_z$ dependence of the peak position in the momentum distribution curves (MDCs) [Figs.~\ref{kz}(c) and\ref{kz}(f)]. The weak $k_z$ dispersion found here is consistent with the recent angular dependent resistivity measurements, which show that the anisotropy of this new superconductor is as large as 30$\thicksim$45 \cite{WHH,NdOsinglecrystal}.

The comprehensive and systematic data presented here allow us to estimate the number of charge carriers in this system based on the Luttinger theorem. We found that its carrier doping level is about \emph{x}=0.16$\pm0.02$, which deviates a lot from \emph{x}=0.5, the  expected value based on the nominal composition. Our band theory calculations  also show that the calculated Fermi surface for \emph{x}=0.16  could well match the experimentally determined Fermi surface [Figs.~\ref{theory}(a) and \ref{theory}(b)].  The possible reasons will be discussed later. On the other hand, we examined the electron correlation in this system. By extracting the peak positions from the MDCs, we obtained the  dispersions for both the $\alpha_{1}$  and $\alpha_{2}$ bands, which are marked by the white open circles in Figs.~\ref{theory}(c) and \ref{theory}(d). Remarkably, the calculated band dispersions for \emph{x}=0.16 (red lines) well coincide with the experimental data along both the $\Gamma$~-~X and X~-~M directions. Therefore, the band renormalization factor $Z$ is about 1 for this compound. Such a weak electron correlation is  consistent with the spatially extended nature of the Bi 6$p$ states that compose the bands near $E_F$. Also, we note that the DFT calculations were performed without the spin-orbit coupling. The remarkable coincidence of the ARPES data with the calculations implies the spin-orbit coupling in this compound is negligible.

 \section{Discussions}

The discrepancy between the nominal   and the actual electron doping levels  may be explained by the non-stoichiometry of the samples, which has been observed both in  our EDS measurements and in others \cite{WHH}. In our samples, bismuth is about 15$\%$ deficient, which directly causes eletron deficiency.
 Quantitatively, one can derive the electron doping level to be around \emph{x}=0.24 from the chemical formula determined by EDS. Considering the error bars of the EDS measurements on light elements, this value is qualitatively consistent with the \emph{x}=0.16$\pm0.02$ experimental value determined from the photoemission data. The fact that the superconductivity could even subsist in the BiS$_2$ planes  with such a high concentration of the bismuth vacancies  implies that  it  is likely an $s$-wave one.

We   note that, for LaO$_{1-x}$F$_{x}$BiS$_{2}$, the phase diagram of $T_C$ versus the fluorine doping level \emph{x} has been reported, where the maximal $T_C$ is around 10K at \emph{x}=0.5\cite{DOSexperimental}. It is possible that the samples studied by us are actually still in the underdoped region in the phase diagram. The $T_C$ might be increased if more electrons could be doped into the system, either by increasing the fluorine doping or by decreasing the bismuth deficiency.  Nevertheless, our results  pose a strong challenge to some of the current theoretical explanations for the superconductivity in BiS$_2$-based compounds.  Without considering the bismuth deficiency of the sample, the tight-binding model or DFT calculations predicted a large and quisi-one-dimensional Fermi surface as shown in Fig.~\ref{theory}(b) for \emph{x}~=0.50 \cite{TBT,Awana,RPA}. Such a Fermi surface results in a strong nesting which would enhance the spin or charge fluctuation in this system.  The pairing mechanism was thus proposed to be similar to that in iron-based superconductors. However, our results show that the superconductivity can survive for a system with a much smaller Fermi surface volume than that predicted by theories. There is  no apparent large Q nesting vector as well. Particularly, we found that the electron correlation is very  weak. Therefore, the unconventional quantum-fluctuation-mediated pairing mechanism is not likely applicable in the BiS$_2$-based superconductors. Instead,  this compound is likely a multi-band superconductor due to   e-ph coupling, such as MgB$_2$ \cite{Mazin,MgB2}.


\section{Conclusion}

To summarize, we have systematically studied the electronic structure of single crystalline NdO$_{0.5}$F$_{0.5}$BiS$_{2}$ by high resolution angle-resolved photoemission spectroscopy. There are two Bi 6$p$ states derived bands around the X point with little $k_z$ dependence, consistent with the theoretical calculations. The band renormalization factor is about 1 in this system, indicating its rather weak electron correlations. Moreover, we found that the actual electron doping in this compound is much smaller than the value expected from the nominal composition, which is likely due to the bismuth dificiency. This gives  much smaller Fermi pockets than those predicted by theoretical calculations for nominal compositions. The small Fermi pocket size and the weak electron correlation found here suggest that the BiS$_2$-based superconductors could be conventional BCS superconductors mediated by e-ph coupling.

\section{ACKNOWLEDGMENTS}

We gratefully acknowledge the experimental support by Dr. M. Shi at SLS and the helpful discussions with Prof. K. Kuroki, Prof. Ang Li, and Prof. Mao Ye. This work was supported by National Basic Research Program of China (973 Program) under the grant Nos. 2011CBA00106, 2011CBA00112, 2012CB921400, and the National Science Foundation of China under Grant No. 11104304. H. F. Yang and D. W. Shen are also supported by the "Strategic Priority Research Program (B)" of the Chinese Academy of Sciences (Grant No. XDB04040300).


\begin{references}

\bibitem {BiOSO} Y. Mizuguchi \emph{et al.}, Phys. Rev. B. \textbf{86}, 220510(R) (2012).

\bibitem {LaO} Y. Mizuguchi \emph{et al.}, J. Phys. Soc. Jpn. \textbf{81}, 114725 (2012).

\bibitem {NdO} S. Demura \emph{et al.}, J. Phys. Soc. Jpn. \textbf{82}, 033708 (2013).

\bibitem{PrO} R. Jha, S. Kumar Singh, and V. P. S. Awana, J. Supercond. Novel Magn. \textbf{26}, 499 (2013).

\bibitem{CeO} J. Xing, S. Li, X. Ding, H. Yang, and H.-H. Wen, Phys. Rev. B \textbf{86}, 214518 (2012).

\bibitem{YbO} D. Yazici \emph{et al.}, Philos. Mag. \textbf{93(6)}, 673 (2013).

\bibitem{TBT} U. Hidetomo, S Katsuhiro, and K Kazuhiko, Phys. Rev. B. \textbf{86}, 220501(R) (2012).

\bibitem{Awana} V. P. S.Awana \emph{et al.}, Solid State Commun. \textbf{157}, 21 (2013).

\bibitem{RPA} G. B. Martins, A. Moreo, and E. Dagotto, Phys. Rev. B \textbf{87}, 081102(R) (2013).

\bibitem{CDW} T. Yildirim, Phys. Rev. B \textbf{87}, 020506(R) (2013).

\bibitem{Xiangang} X. Wan, H.-C. Ding, S. Y. Savrasov, and C.-G. Duan, Phys. Rev. B \textbf{87}, 115124 (2013).

\bibitem{PhononLi} B. Li, Z. W. Xing, and G. Q. Huang, Europhys. Lett. \textbf{101}, 47002 (2013).

\bibitem{ZDW} T. Zhou and Z. D. Wang, J. SUPERCOND. NOV. MAGN. \textbf{26}, 2735 (2013).

\bibitem{penetration} Shruti, P. Srivastava and S. Patnaik, J. Phys. Condens. Matter \textbf{25}, 31 (2013).

\bibitem{Morenzoni} P. K. Biswas \emph{et al.}, arXiv:1309.7282.

\bibitem{Putti} G. Lamura \emph{et al.}, arXiv:1311.0457.

\bibitem{Neutron} J. Lee \emph{et al.}, Phys. Rev. B \textbf{87}, 205134 (2013).

\bibitem{WHH} J.  Liu~\emph{et al.}, arXiv:1310.0377.

\bibitem{LXYang} L. X. Yang \emph{et al.}, Phys. Rev. B\textbf{ 82}, 104519 (2010).

\bibitem{Hall} S. K. Singh~\emph{et al.}, J. Am. Chem. Soc. \textbf{134}, 16504 (2012).

\bibitem{Hall2} S. Li~\emph{et al.}, Sci China-Phys Mech Astron \textbf{56}, 2019 (2013).

\bibitem{Hufner}  S. H¨¹fner, Photoelectron Spectroscopy, Third edition (Springer-Verlag, Berlin, 2003).



\bibitem{NdOsinglecrystal} M.  Nagao~\emph{et al.}, J. Phys. Soc. Jpn. \textbf{82}, 113701 (2013).


\bibitem{DOSexperimental} K. Deguchi~\emph{et al.}, EPL, \textbf{101}, 17004 (2013).

\bibitem{Mazin} I. I. Mazina and V. P. Antropov, Physica C: Superconductivity \textbf{385}, 49 (2002).

\bibitem{MgB2} S. Souma~\emph{et al.}, Nature \textbf{423}, 65 (2003).



\end{references}
\end{document}